\documentclass[iop]{emulateapj}
\usepackage{apjfonts}
\usepackage{epsf}
\usepackage{xcolor}
\usepackage{longtable}

\begin{document}
\slugcomment{}
\shortauthors{Morales et al.}
\shorttitle{Mrk 817 X-ray/UV Correlation}

\title{X-ray and UV Monitoring of the Seyfert 1.5 Galaxy Markarian 817}

\author{Anthony~M.~Morales\altaffilmark{1},
Jon~M.~Miller\altaffilmark{1},
Edward~M.~Cackett\altaffilmark{2},
Mark~T.~Reynolds\altaffilmark{1},
and
Abderahmen~Zoghbi\altaffilmark{1}
}

\altaffiltext{1}{Department of Astronomy, University of Michigan, 1085
  South University Avenue, Ann Arbor, MI 48109-1107, USA,
  ammoral@umich.edu}
\altaffiltext{2}{Department of Physics \& Astronomy, Wayne State
  University, 666 W. Hancock Street, Detroit, MI, 48201, USA}
  
\begin{abstract}
We report the results of long-term simultaneous X-ray and UV
monitoring of the nearby ($z=0.03145$) Seyfert 1.5 galaxy Mrk 817
using the Neil Gehrels \textit{Swift} Observatory XRT and UVOT.  Prior
work has revealed that the X-ray flux from Mrk 817 has increased by a
factor of ~40 over the last 40 years, whereas the UV emission has
changed by a factor of ~2.3.  The X-ray emission of Mrk 817 now
compares to some of the brightest Seyferts, but it has been poorly
studied in comparison. We find that the X-ray (0.3-10.0 keV) and the
UVM2 (roughly 2000--2500\AA) fluxes have fractional variability
amplitudes of $0.35$ and $0.18$, respectively, over the entire
monitoring period (2017 Jan. 2 to 2018 Apr. 20).  A
cross-correlation analysis is performed on the X-ray (0.3-10.0 keV)
and UVM2 light curves over the entire monitoring period, a period of
less frequent monitoring (2017 Jan. 2 to 2017 Dec. 11), and a period of 
more frequent monitoring (2018 Jan. 12 to 2018 Apr. 20). The analysis reveals
no significant correlation between the two at any given lag for all
monitoring periods. Especially given that 
reverberation studies have found significant lags between optical/UV 
continuum bands and broad optical lines in Mrk 817, the lack of a 
significant X-ray--UV correlation may point to additional 
complexities in the inner or intermediate disk. Mechanical (e.g.,a
funnel in the inner disk) and/or relativistic beaming of the X-ray
emission could potentially explain the lack of a correlation.
Alternatively, scattering in an equatorial wind could also diminish
the ability of more isotropic X-ray emission to heat the disk itself.
\end{abstract}


\section{Introduction}
Active galactic nuclei (AGN) are powered by accretion onto a
supermassive black hole.  At high fractions of the Eddington limit,
this process is mediated by an accretion disk that peaks in the UV.  A
``corona'' is also inferred through X-ray emission in many AGN.
Though the corona and accretion disk may be physically distinct, the
X-ray and UV emission from these regions may interact with each other.
On long timescales (weeks to months), variations in the mass accretion rate through the
disk should cause changes in the rate of viscous dissipation, and
thereby the UV flux; Compton up-scattering in the corona may then
drive variations in the X-ray flux.  On shorter timescales (less than 1-2 days), 
rapid variability in the corona -- perhaps due to magnetic activity -- can
drive X-ray flux changes that are reprocessed into UV emission in the
disk.  On both long and short timescales, then, a very simple picture
predicts clear lags and correlations.

Recent research into X-ray and UV variability has produced mixed
results.  Several teams looking at different sources have found cases
where the X-ray emission leads the UV emission (e.g. McHardy et
al. 2014; Troyer et al. 2016; Edelson et al. 2015; Edelson et al. 2017; 
McHardy et al. 2018; Pal and Naik 2018),
consistent with X-ray reprocessing.  These lags are longer than
predicted, though, and they are attributed to larger accretion disks
or extra reprocessing mechanisms.  In the case of NGC 5548, the 
X-ray emission has been found to lead the optical emission on 
timescales of ~1 year (Uttley et al. 2003), but it has been noted recently
that the X-ray and UV/optical emissions correlate poorly on shorter 
timescales, which again require extra reprocessing mechanisms to 
explain such results
(Edelson et al. 2015; Gardner \& Done 2017; Starkey et al. 2017). 
Others, however, have found no
correlation between the X-ray and UV emission, attributing this to
light bending of a centrally compact X-ray source near to the black
hole, X-ray flux in our line of sight not being correlated with the
X-ray emission that is reprocessed in the disk, and/or other
mechanisms driving the observed UV variability (Robertson et al. 2015;
Buisson et al. 2018).

It has also been previously noted that the appearance of significant
lags can change depending on the interval or time scale being studied.
For example, Gallo et al. (2018) examined X-ray and UV light curves of
Mrk 335 and found no correlation between the two, over an eleven year
span.  However, when they restricted their analysis to only the 8th
year of monitoring, they found a highly significant positive
correlation of the X-ray leading the UV.  They attributed this
correlation to a giant X-ray flare that occurred during their
monitoring.  This and the examples above reveal the complexity of 
X-ray/UV radiation and the need for a larger number of multi-wavelength
studies to better understand disks and disk--corona connections in AGN. 

In this paper, we examine the relationship between X-ray and UV
variability of the nearby ($z=0.03145$) Seyfert 1.5 galaxy Mrk 817
($\log(M_{BH}/M_{\odot})=7.586$; Bentz \& Katz 2015).  The X-ray
emission of Mrk 817 has changed drastically since its discovery
decades ago.  Previously, the modest X-ray flux of Mrk 817 made it
largely inaccessible.  However, by 2009, its X-ray flux had increased
by a factor of $\sim40$ since 1990 (Winter et al. 2011), making Mrk
817 one of the brightest Seyferts in the X-ray regime.  In contrast,
the UV emission only varied by a factor of $\sim2.3$ over a 30 year
period (Winter et al. 2011).  The fact that the X-ray light curve has previously shown
marked variability while the UV light curve does not, makes Mrk 817 a
particularly intriguing AGN.  Yet, in comparison to other X-ray-bright
Seyfert galaxies, Mrk 817 has been overlooked.

\begin{figure*}[htb!]
	\centering
	\includegraphics[scale=1]{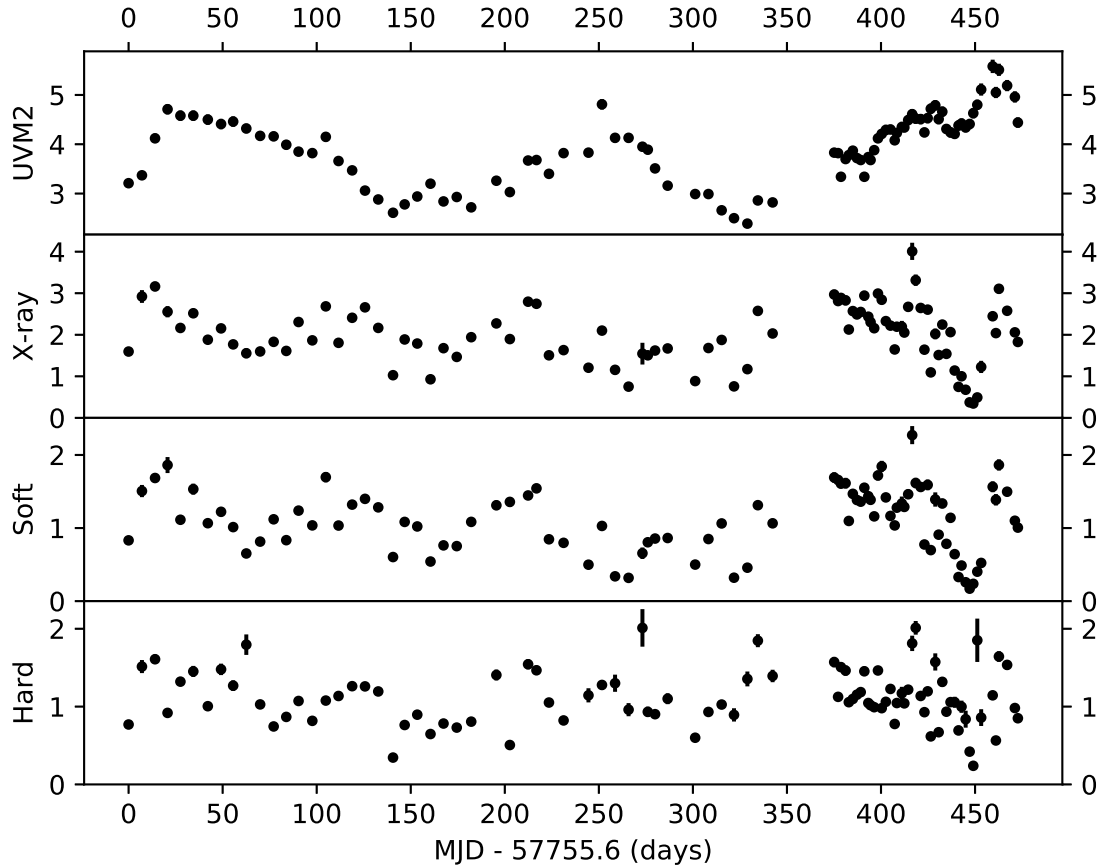}
	\figcaption{The {\it Swift} UVOT/UVM2 flux density
          $(10^{-14}~\text{erg}~\text{s}^{-1}~\text{cm}^{-2}~\text{\AA}^{-1})$,
          XRT X-ray (0.3-10.0 keV) flux
          $(10^{-11}~\text{erg}~\text{s}^{-1}~\text{cm}^{-2})$, XRT
          soft X-ray (0.3-2.0 keV) flux
          $(10^{-11}~\text{erg}~\text{s}^{-1}~\text{cm}^{-2})$, and
          XRT hard X-ray (2.0-10.0 keV) flux
          $(10^{-11}~\text{erg}~\text{s}^{-1}~\text{cm}^{-2})$ of Mrk
          817 are shown from top to bottom, respectively.  
          Most error bars are too small to be visible.  The
          UVM2 flux density shown is the corrected flux density, which
          was calculated using \textsc{uvotsource}. The X-ray
          (0.3-10.0 keV) flux is calculated using a power-law model in
          \textsc{xspec}, while the soft and hard X-ray fluxes are
          calculated using a broken power-law model with break energy
          2.0 keV in \textsc{xspec}. See Section 2 for more details.}
	\label{fig:fig1}
\end{figure*}

The most recent X-ray spectral study of Mrk 817, by Winter et
al. (2011), used the six X-ray observations from {\it Swift} and
single {\it XMM-Newton} exposure that were available at the time.
They found that the optical/UV fluxes were similar and the
optical-X-ray SED changes between observations were due to variations
in the X-ray luminosity/spectral shape, concluding that the UV and
X-ray are not correlated.  Winter et al.\ (2011) also reported a
strong positive correlation between the X-ray spectral slope and X-ray
luminosity.  We re-examine these results more thoroughly using a
cross-correlation analysis on 94 new X-ray and UV observations from
{\it Swift} .  In Section 2, we describe the observations and data
reduction.  Section 3 details and provides results of a
cross-correlation analysis of the X-ray and UV light curves.  We end
with a discussion of our results in Section 4.

\section{Observations and Data Reduction}
Mrk 817 was monitored simultaneously using the \textit{Swift} XRT
(Burrows et al.\ 2005) and the \textit{Swift} UVOT (Roming et
al.\ 2005) from 2017 Jan. 2 to 2018 Apr. 20.  
Over this $473$-day period,
94 observations were considered, separated in time by an average of
$\sim5.1$ days.

The XRT exposures were taken in ``photon counting'' mode with a
$23.6\times23.6$ arcmin FOV.  Source events were extracted from a
circle with a radius 20 pixels, centered on coordinates provided in
NED\footnote{The NASA/IPAC Extragalactic Database (NED) is operated by
  the Jet Propulsion Laboratory, California Institute of Technology,
  under contract with the National Aeronautics and Space
  Administration.}.  Background events were extracted using a circular
region of radius 20 pixels, off-center from the source coordinates and
outside the source region.  Ancillary response files were created for
each observation using \textsc{xrtmkarf}, each of which included the
provided exposure map to account for bad CCD pixels and columns. Using
\textsc{grppha}, the spectra were grouped such that there are at least
10 counts in each bin.

The extracted spectra were then fitted with a power-law model over the 
0.3-10.0 keV energy range (mean count rate of 
$0.5466\pm0.0025~\text{cts}~\text{s}^{-1}$) using \textsc{xspec v.12.9.1p}.  An absorbed 
power-law model is not considered, as the galactic HI column density is too small
($\text{N}_{\text{H}}=1.50\times10^{20}~\text{cm}^{-2}$, Dickey \& Lockman 1990) to
produce any discernible effects within the short exposures.  For each spectrum, 
the photon index and normalization, and the corresponding $1\sigma$ errors on these
parameters, were recorded.  The flux in the fitting band was obtained
using the ``flux'' command within XSPEC; flux errors were calculated
by assuming that the flux has the same fractional error as the
power-law normalization.  We find that this gives a more conservative
estimation of error than the one provided by \textsc{xspec}. The X-ray
light curve can be seen in the panel labeled ``X-ray'' of Figure 1. 

We separately modeled all X-ray spectra using a broken power-law
model, in an attempt to allow for a physically distinct ``soft
excess''.  The break energy was fixed at 2 keV in each observation,
and the flux for the soft (0.3--2.0~keV) and hard (2.0--10.0~keV)
X-rays were recorded.  The flux errors were calculated in the same
manner as before.  The soft and hard X-ray light curves are shown in
the panels labeled ``Soft'' and ``Hard'' of Figure 1, respectively.

The UV images were taken with the UVOT/UVM2 filter, which has the
smallest red-leak of the {\it Swift} UVOT UV filters.  The UVM2 filter
has a central wavelength of 2246 \AA, an effective wavelength of 2231
\AA, and a FWHM of 498 \AA\ (Poole et al. 2008).  Source events were
extracted from the images using a circular region of radius 5
arc-seconds centered on the source.  Background emission was accounted
for using an annulus (33 arc-second inner-radius, 43 arc-second
outer-radius) centered on the the source.  We calculated the flux
density of the UVM2 images using the \textsc{ftools} function
\textsc{uvotsource}.  It should be noted that there can be coincidence
losses due to the brightness of the source, and \textsc{uvotsource}
provides a reconstructed ``true'' flux density; this is the flux
density we recorded for each observation.  The error on each flux
density value was calculated as
$\sqrt{\text{stat-err}^2+\text{sys-err}^2},$ where stat-err and
sys-err are the statistical error and systematic error on the flux
density, respectively.  The light curve for the UVM2 is given in the
panel labeled ``UVM2'' of Figure 1.  The mean corrected count rate for the 
entire monitoring period is $46.66\pm0.11~\text{cts}~\text{s}^{-1}$ with
a mean corrected count rate factor of $1.627\pm0.004$.
It is well-known that UV ``dropouts" can 
occur due to suspect regions of the detector 
(e.g. Edelson et al. 2015; Edelson et al. 2017). We do not account 
for ``dropouts" here, as the UVM2 light curve has very few obvious points
that demonstrate such features. 

Light curve data and exposure times for each of the 94 
observations is provided in Table 1.

\section{Analysis \& Results}

\subsection{Photon Index/X-ray Relationship}
A plot of the power-law photon index and corresponding X-ray (0.3-10.0
keV) flux is provided in Figure 2.  The mean power-law photon 
index is $2.0047\pm0.0117$.  
We calculated a Spearman
correlation coefficient of $r=0.34$ with a two-sided p-value of
$p=9.6\times10^{-4}$.  In contrast, Winter et al. (2011) found a
positive correlation between the photon index and X-ray luminosity,
with a Spearman correlation coefficient of $r=0.77$ and a two-sided
p-value of $p=0.05$.  Our analysis includes a much larger set of
observations, and may therefore more accurately reflect the degree to
which the power-law properties are correlated.  Combining our points
with the data presented in Winter et al. (2011) slightly increases the
correlation we originally calculated: $r=0.38$ with
$p=1.1\times10^{-4}$.

\begin{figure}[htb!]
	\center
	\includegraphics[width=9cm]{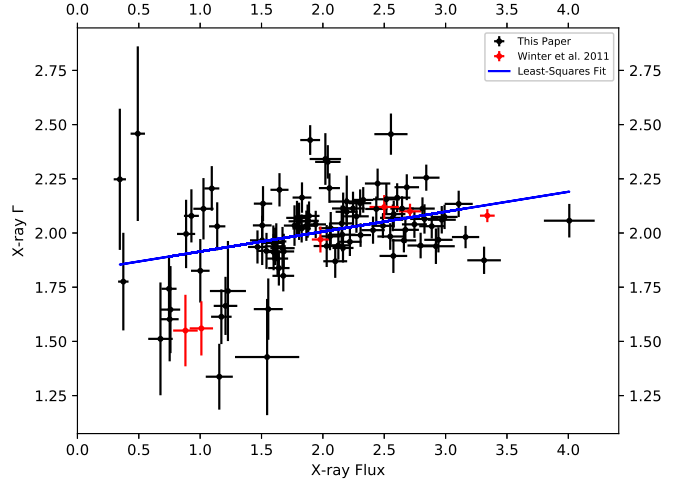}
	\caption{The power-law photon index versus
          the X-ray (0.3-10.0 keV) flux
          $(10^{-11}~\text{erg}~\text{s}^{-1}~\text{cm}^{-2})$.  The
          data first presented in our analysis are shown in black;
          data reported by Winter et al. (2011) are shown in red.  The
          plot has a Spearman correlation coefficient of $r=0.34$ with
          a two-sided p-value of $p=9.6\times10^{-4}$.  The mean 
          power-law photon index is $2.0047\pm0.0117$.  The least-squares 
          linear fit $y=mx+b$ gives $m=0.092,b=1.823$ with squared 
          correlation coefficient $R^2=0.111$. See Section 3.1 for more
          details.}
\end{figure}

\begin{figure}[htb!]
	\center
	\includegraphics[width=9cm]{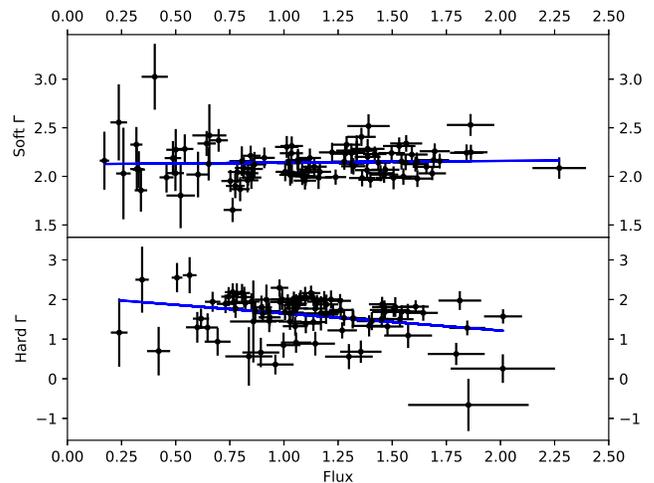}
	\caption{{\it Top Pannel}: The broken power-law {\em soft} photon index
          versus {\em soft} X-ray flux
          $(10^{-11}~\text{erg}~\text{s}^{-1}~\text{cm}^{-2})$.  The
          data give a Spearman correlation coefficient of $r=0.12$
          with a two-sided p-value of $p=0.24$.  The mean {\em soft} 
          photon index is $2.143\pm0.015$.  The least-squares 
          linear fit gives $m=0.017,b=2.124$ with squared 
          correlation coefficient $R^2=0.042$.  {\it Bottom Panel}:
          The broken power-law {\em hard} photon index versus the {\em
            hard} X-ray flux
          $(10^{-11}~\text{erg}~\text{s}^{-1}~\text{cm}^{-2})$.  The
          data give a Spearman correlation coefficient of $r=0.25$
          with a two-sided p-value of $p=0.013$. The mean {\em hard} 
          photon index is $1.601\pm0.032$.  The least-squares 
          linear fit gives $m=-0.433,b=2.082$ with squared 
          correlation coefficient $R^2=0.086$.  See Section 3.1 for more
          details.}
	\label{fig:fig3}
\end{figure}

\begin{figure}[htb!]
	\center
	\includegraphics[width=9cm]{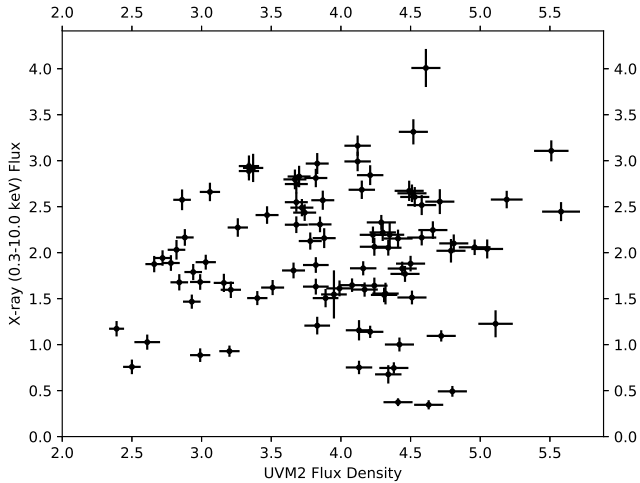}
	\caption{The X-ray (0.3-10.0 keV) flux
          $(10^{-11}~\text{erg}~\text{s}^{-1}~\text{cm}^{-2})$ versus
          the UVM2 flux density
          $(10^{-14}~\text{erg}~\text{s}^{-1}~\text{cm}^{-2}~\text{\AA}^{-1})$.
          We calculate a Pearson correlation coefficient of $r=0.11$
          with two-sided p-value of $p=0.29$.  See Section 3.2 for
          more details.}
	\label{fig:fig4}
\end{figure}

\begin{figure*}[htb!]
	\center
	\includegraphics[scale=0.96]{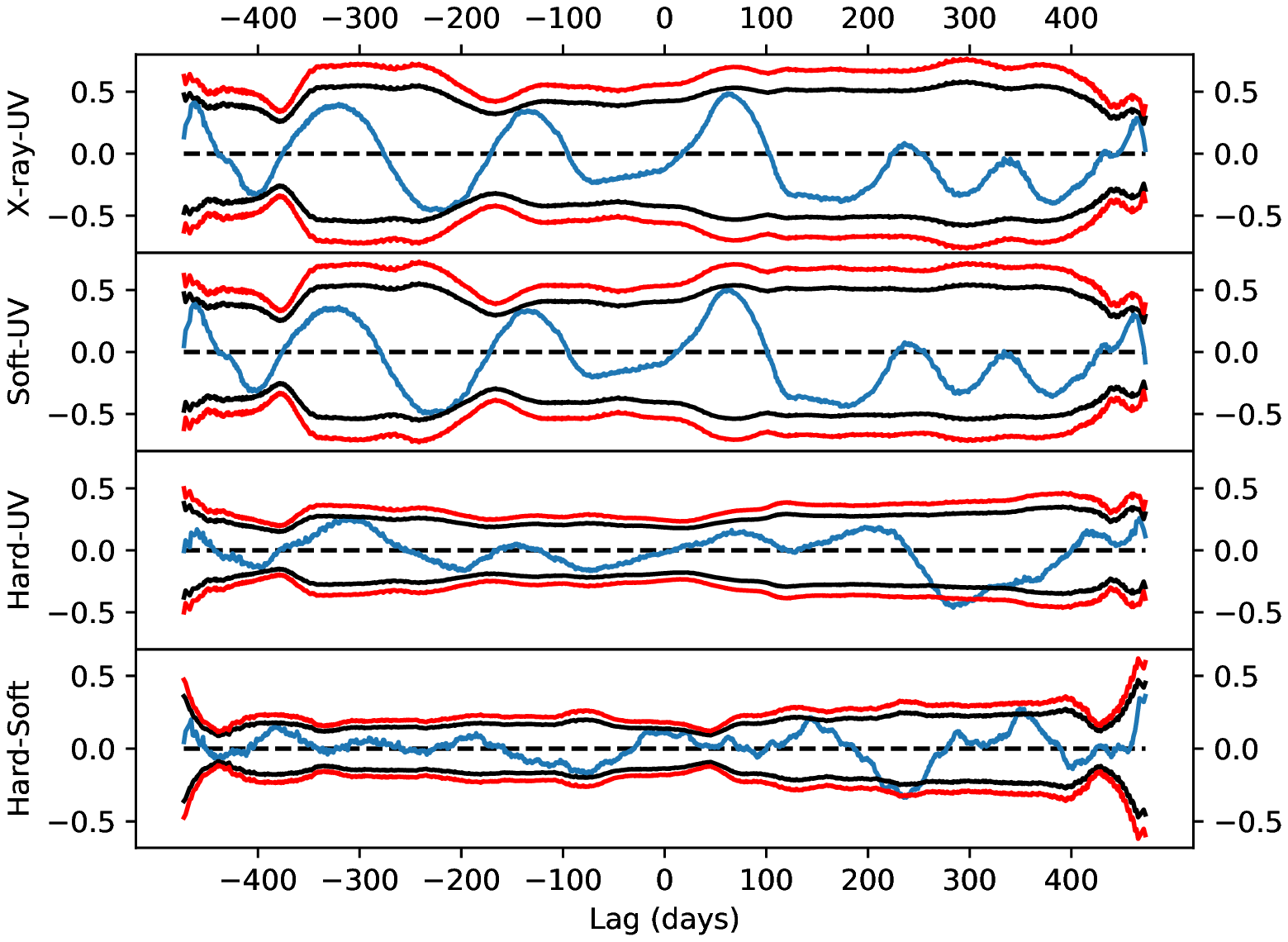}
	\caption{The various calculated DCFs. The black and red curves
          represent 95 and 99 percent confidence curves, respectively.
          There is a lack of correlation in the X-ray-UV DCF and the
          soft X-ray-UV DCF, as all values are less than 95\%
          confident. A 99\% significant anti-correlation exists in the
          hard X-ray-UV DCF at a lag of $\sim284$ days, corresponding
          to the hard X-ray flux leading the UV flux by $\sim284$
          days.  The DCF between the hard and soft X-rays also has a
          $99\%$ significant anti- correlation at $\sim235$ days,
          corresponding to the hard X-ray flux leading the soft X-ray
          flux.  See Section 3.2 for calculation details and Section 4
          for a discussion of the results.}
	\label{fig:fig5}
\end{figure*}

We also provide plots for the soft X-ray photon index versus the soft
X-ray flux and the hard photon index versus the hard X-ray flux in
Figure 3.  The soft photon index and hard photon index
correspond to the 0.3--2.0~keV and 2.0--10.0~keV energy ranges of the
broken power-law model, respectively.  The mean soft photon index and 
mean hard photon index are $2.142\pm0.015$ and $1.601\pm0.032$, 
respectively.  The soft photon index and soft
X-ray flux give a Spearman correlation coefficient of $r=0.12$ with
two-sided p-value of $p=0.24$.  The hard photon index and hard X-ray
flux produce a Spearman correlation coefficient of $r=0.25$ with
two-sided p-value of $p=0.013$.

\subsection{X-ray/UV Correlation Analysis}

A plot (Figure 4) of the X-ray (0.3-10.0 keV) flux versus
the UVM2 flux shows minimal correlation between the two bands.
Indeed, an analysis of the two produces a Pearson correlation
coefficient of $r=0.11$ with a two-sided p-value of $p=0.29$.

A cross-correlation analysis was performed in order to test for
correlations at different lags.  We use the discrete correlation
function (DCF; Edelson \& Krolik 1988, Robertson et al. 2015), because
of the uneven sampling of the X-ray and UV observations.  A positive
value on the lag axis corresponds to the X-ray flux leading the UV
flux.  The DCF requires the lag time axis to be separated into bins of
equal width, $\delta\tau$, which is a free variable.  By the
definition of the DCF, these bins must contain at least one data
point, and -- if errors are to be calculated -- each bin must contain
at least two data points.  Choosing a value for $\delta\tau$ is a
tradeoff between statistical accuracy and resolution; the former
requires larger $\delta\tau$ and the latter requires smaller
$\delta\tau$.  We set $\delta\tau=50$ days, as this allowed for at
least 20 points in each bin, providing reasonable statistical accuracy
and a DCF with discernible details.

We calculated 95 and 99 percent confidence curves to evaluate the
significance of various potential lags.  We first simulated 1000 X-ray
light curves using the method of Timmer \& K\"onig (1995).  This
procedure requires knowledge of the power spectral density (PSD),
which we estimate to be a power-law with a slope of -1.1 and a
normalization value of 0.011; this estimation follows the method of
Zoghbi et al. (2013).  Values are then selected from each simulated
light curve such that they correspond to the times at which the real
X-ray observations were recorded.  The DCF between these simulated
values and the real UV light curve is calculated.  This produces a
distribution of 1000 correlation values at each lag, which we then used
to calculate 95 and 99 percent confidence values.

We report no correlation above a 95\% confidence value at any given
lag, as can be seen in the first panel of Figure 5.
However, via visual inspection, there appears to be an
anti-correlation between the X-ray and the UV light curves from
2018 Jan. 12 to 2018 Apr. 20 (MJD 58130 to MJD 58228), a time period of
approximately 98 days in which the observation cadence increases.  The
DCF plot shows an anti-correlation at a lag of $\sim18.3$ days (see
Figure 6), significant at the 95\% level of confidence,
corresponding to the X-ray flux leading the UV flux.  Owing to this
modest significance, the lack of a clear physical interpretation for an 
anti-correlation, and the robust nature of the DCF\footnote{This robust 
nature refers to the method to calculate the DCF as an approximation to 
a true cross-correlation function.}, 
this putative signature must be regarded skeptically.  The light curves
were cross-correlated for a period of less frequent observation
(2017 Jan. 2 to 2017 Dec. 122, MJD 57755 to MJD 58098), and no significant
correlation was found.

\begin{figure}[htb!]
	\center
	\includegraphics[width=8.5cm]{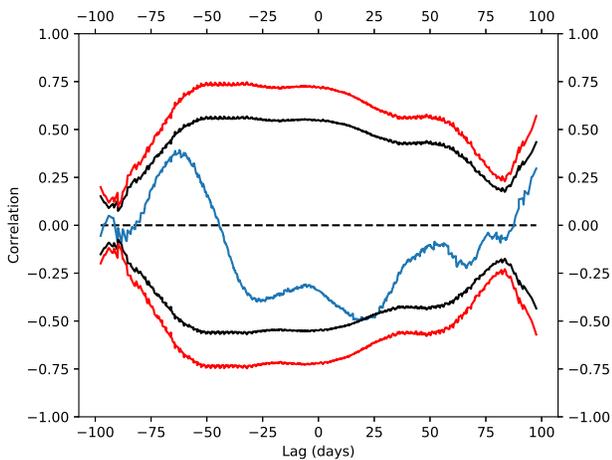}
	\caption{The DCF between the higher cadence portions of the
          X-ray (0.3-10.0 keV) and UVM2 light curves is shown.  A 95\%
          significant anti-correlation exists at a lag of $\sim18.3$ days,
          corresponding to the X-ray leading the UV. See Section 3.2
          for more details.}
	\label{fig:fig6}
\end{figure}

The soft and hard X-ray bands were each cross-correlated with the UVM2
as well; these DCFs are plotted in the second and third panels of
Figure 5, respectively.  The soft X-ray band produces a DCF
similar to the full X-ray band (0.3-10.0 keV), showing no
correlation.  The DCF of the hard X-ray band with the UVM2, however,
produces a 99\% confident anti-correlation at a lag of $\sim284$ days,
meaning the X-ray leads the UV by $\sim284$ days.  The fact that the
lag time scale is comparable to the period over which the monitoring
was conducted indicates that this signal should be regarded cautiously. 
Horne et al. (2004) suggest that the length of the monitoring period should 
be at least 3 times longer than the lags of interest, whereas Welsh (1999) 
recommends a monitoring period of at least 4 times longer, but preferably 
$\sim10$ times longer, than the lags of interest.  In this paper, we follow 
the suggestion of Horne et al. (2004). 

A cross-correlation of the hard and soft X-ray light curves reveals a $99\%$
significant anti-correlation at a lag of $\sim235$ days, corresponding to
the hard X-ray flux leading the soft X-ray flux (see the fourth panel
of Figure 5).  There also exists $95\%$ confident positive
correlations at lags of $\sim144$ days and $\sim349$ days.  These
positive correlations may also be spurious: 
the lags are not highly significant, and the robust nature of the
DCF can produce such results.  Here again, the longest lags are 
also close to the full period of over which the monitoring was conducted.
Visually inspecting the soft and hard X-ray light curves, it appears that 
the two light curves are tracking each other. We find that the hard and 
soft X-ray light curves are indeed correlated at lag $\tau=0$ by calculating a 
Pearson correlation coefficient of $r=0.464$ with a two-sided p-value of 
$p=2.49\times10^{-6}$, corresponding to a $>99\%$ significance.  
The disagreement at $\tau=0$ between the DCF and the Pearson 
coefficient is due to the binning requirement of the DCF. 
 
\section{Discussion}
We have undertaken an extensive X-ray and UV monitoring campaign to
better understand Mrk 817, a bright, nearby, Seyfert-1.5 AGN.
Although the source is now as bright as the best--known Seyferts, it
was much fainter in past decades, and as a result it has not been
studied extensively.  We cross-correlated the X-ray and UV light
curves using the standard DCF; separate cross-correlations were
undertaken for the soft and hard X-ray flux, in recognition of
potentially distinct components within the spectrum.  We do not find
any lag signals at a high level of statistical significance.  In this
section, we offer plausible physical reasons why strong
correlations might be absent.

Many similar analyses have found significant short lags, with the
X-rays leading the UV in a manner broadly consistent with reprocessing
(McHardy et al. 2014; Troyer et al. 2016; Edelson et al. 2015; 
Edelson et al. 2017; McHardy et al. 2018; Pal and Naik 2018).  In some 
cases, the observed lags are slightly longer than
anticipated by standard thin disk models; this may suggest that AGN
disks do not follow such prescriptions or the lags are contaminated by 
the continuum emission from the broad-line region 
(Korista \& Goad 2001; Cackett et al. 2018).  
In the case of Mrk 817, however, the DCFs that we calculated find no 
robust correlations above the 95\% confidence between the X-ray 
(0.3-10.0 keV) and UV light curves.  Putative lag signals on long time scales 
-- noted above -- are too close to the duration of the monitoring to be credible.
Though uncommon, non-correlations have been reported recently
(Robertson et al. 2015; Buisson et al. 2018). This lack of a significant 
X-ray/UV correlation is inconsistent with the optical variability of Mrk 817, as it 
has been reverberation mapped, i.e. lags have been calculated 
between the H$\beta$ and continuum light curves 
(Peterson et al 1998; Denney et al. 2010). This suggests that, 
in this source, the X-ray variability is not the primary driver of 
photoionization and line variability in the BLR.  This is at odds with 
the standard lamppost model whereby a small, centrally located X-ray 
source drives variability in the UV/optical continuum and emission lines.

Geometric beaming -- or perhaps ``funneling'' -- of the X-ray emission
represents one potential means by which reprocessing of X-ray emission into
UV flux might be prevented.  For an AGN with a high-Eddington fraction,
for instance, the inner accretion disk may fail
to cool efficiently, giving it an increased scale height above the
local surface of the disk.  This geometry may serve as
an inner funnel that could block the X-ray emission from larger disk radii.  
This explanation requires that we have a privileged viewing angle into the
funnel.  Moreover, Mrk 817 has a low-Eddington fraction:
$L_{\text{bol}}/L_{\text{Edd}}=0.1950$ ($\log L_{\text{bol}}=44.99$,
Woo \& Urry 2002).  Recent work on NGC 5548 and NGC 4151,
in particular, has suggested that additional geometries 
may be needed to account for X-ray and UV correlations 
(e.g., Edelson et al.\ 2017; Gardner \& Done 2017).  Moreover, 
X-ray observations of NGC 4151 find evidence of a warp or bulge 
in the inner accretion disk (Miller et al.\ 2018; Zoghbi et al.\ 2018) though 
its Eddington fraction is even lower than that of Mrk 817.

A qualitatively distinct but physically similar scenario is that the
X-ray flux is relativistically beamed away from the disk, owing to
bulk motion normal to the disk.  If the X-ray corona is the base of a
jet (Markoff, Nowak, \& Wilms 2005; Miller et al. 2006), where 
acceleration away from the disk may occur, then the
velocities required to escape the regions closest to the black hole
may serve to naturally beam the X-ray emission away from the disk.
Like most other Seyferts, however, Mrk 817 is not radio--loud, so if
this process is at work the corona might be the base of a ``failed''
jet.

It is also possible that a dense outflow from the inner disk could
inhibit the reprocessing of X-rays into UV flux in the disk.  A disk
wind that covers the inner accretion disk might scatter a fraction of
the X-ray emission before this flux reaches the optically-thick
accretion disk.  Winter et al.\ (2011) find no evidence of X-ray
absorption consistent with an outflow; however, disk winds --
especially in the inner disk -- may be primarily equatorial.  If we
observe Mrk 817 at a low inclination (closer to the pole), a wind may
operate but not be detected.  Recent observations of NGC 5548 have
also revealed evidence of dense, transient outflows from the inner
disk (Kaastra et al.\ 2014).  Additional X-ray and UV spectroscopy of
Mrk 817 may reveal if the source was in a similar state during the
period of our monitoring, and/or when Winter et al.\ (2011) studied
the source with {\it XMM-Newton}. Obscuration by transient clouds from 
a warm absorber and a neutral absorber can mask the intrinsic variability
of the X-ray emission, which could lead to a lack of correlation 
between the X-ray and UV emission.

\pagebreak

\begin{longtable}{c|c|c|c|c|c|c|c|c|c|c}
\caption{Data. The observation I.D. (denoted OBSID), start time in MJD, exposure time in seconds, 
UVM2 flux density and flux density error 
($10^{-14}~\text{erg}~\text{s}^{-1}~\text{cm}^{-2}~\text{\AA}^{-1}$), X-ray (0.3-10.0 keV), soft X-ray 
(0.3-2.0 keV), and hard X-ray (2.0-10.0 keV) fluxes and flux 
errors ($10^{-11}~\text{erg}~\text{s}^{-1}~\text{cm}^{-2}$) for each observation are listed. Start time MJD 
57755.6 is 2017 Jan. 02 at 14:24:00.000 UTC, and end time MJD 58228.4 is 2018 Apr. 20 at 
09:36:00.000 UTC.}  \label{tab:table1}\\

\hline \multicolumn{1}{c}{\textbf{OBSID}} & \multicolumn{1}{c}{\textbf{Start Time}} & \multicolumn{1}{c}{\textbf{Exposure Time}} & \multicolumn{1}{c}{\textbf{UVM2}} & \multicolumn{1}{c}{\textbf{UVM2 Error}} & \multicolumn{1}{c}{\textbf{X-ray}} & \multicolumn{1}{c}{\textbf{X-ray Error}} & \multicolumn{1}{c}{\textbf{Soft}} & \multicolumn{1}{c}{\textbf{Soft Error}} & \multicolumn{1}{c}{\textbf{Hard}} & \multicolumn{1}{c}{\textbf{Hard Error}}\\ \hline 
\endfirsthead

\multicolumn{11}{c}%
{{\bfseries \tablename\ \thetable{} ~(continued)}} \\
\hline \multicolumn{1}{c}{\textbf{OBSID}} & \multicolumn{1}{c}{\textbf{Start Time}} & \multicolumn{1}{c}{\textbf{Exposure Time}} & \multicolumn{1}{c}{\textbf{UVM2}} & \multicolumn{1}{c}{\textbf{UVM2 Error}} & \multicolumn{1}{c}{\textbf{X-ray}} & \multicolumn{1}{c}{\textbf{X-ray Error}} & \multicolumn{1}{c}{\textbf{Soft}} & \multicolumn{1}{c}{\textbf{Soft Error}} & \multicolumn{1}{c}{\textbf{Hard}} & \multicolumn{1}{c}{\textbf{Hard Error}}\\ \hline
\endhead

\hline \multicolumn{11}{r}{{Continued on next page...}} \\ \hline
\endfoot

\hline
\hline

\endlastfoot

00037592002  &  57755.6  &  981.4  &  3.21  &  0.073  &  1.596  &  0.0893  &  0.832  &  0.0501  &  0.77  &  0.0464  \\
00037592003  &  57762.7  &  1088.8  &  3.37  &  0.073  &  2.921  &  0.1536  &  1.506  &  0.0835  &  1.514  &  0.0839  \\
00037592004  &  57769.7  &  996.4  &  4.12  &  0.095  &  3.163  &  0.1111  &  1.685  &  0.0636  &  1.608  &  0.0607  \\
00037592005  &  57776.3  &  1001.4  &  4.71  &  0.104  &  2.555  &  0.1352  &  1.862  &  0.1096  &  0.919  &  0.0541  \\
00037592006  &  57783.2  &  1213.7  &  4.58  &  0.104  &  2.164  &  0.0856  &  1.113  &  0.0479  &  1.32  &  0.0568  \\
00037592007  &  57790.0  &  789.2  &  4.58  &  0.104  &  2.517  &  0.1087  &  1.532  &  0.075  &  1.452  &  0.0711  \\
00037592008  &  57797.7  &  986.4  &  4.5  &  0.104  &  1.881  &  0.0847  &  1.066  &  0.0523  &  1.005  &  0.0493  \\
00037592009  &  57804.8  &  949.0  &  4.41  &  0.104  &  2.154  &  0.097  &  1.222  &  0.0608  &  1.478  &  0.0735  \\
00037592010  &  57811.2  &  941.5  &  4.46  &  0.104  &  1.769  &  0.0867  &  1.013  &  0.0542  &  1.269  &  0.0679  \\
00037592011  &  57818.2  &  936.5  &  4.32  &  0.095  &  1.555  &  0.1179  &  0.652  &  0.0478  &  1.796  &  0.1316  \\
00037592012  &  57825.6  &  1071.3  &  4.17  &  0.095  &  1.599  &  0.0774  &  0.815  &  0.042  &  1.028  &  0.053  \\
00037592013  &  57832.7  &  1121.3  &  4.16  &  0.095  &  1.83  &  0.0768  &  1.12  &  0.0513  &  0.745  &  0.0341  \\
00037592014  &  57839.5  &  1013.9  &  3.99  &  0.095  &  1.611  &  0.0861  &  0.834  &  0.0476  &  0.867  &  0.0494  \\
00037592015  &  57846.0  &  1211.2  &  3.85  &  0.082  &  2.307  &  0.0851  &  1.239  &  0.049  &  1.072  &  0.0424  \\
00037592016  &  57853.3  &  981.4  &  3.82  &  0.092  &  1.865  &  0.0853  &  1.036  &  0.0511  &  0.817  &  0.0403  \\
00037592017  &  57860.5  &  983.9  &  4.15  &  0.095  &  2.684  &  0.1  &  1.696  &  0.0697  &  1.077  &  0.0443  \\
00037592018  &  57867.2  &  906.5  &  3.66  &  0.082  &  1.806  &  0.0858  &  1.034  &  0.0535  &  1.136  &  0.0588  \\
00037592019  &  57874.5  &  976.4  &  3.47  &  0.082  &  2.409  &  0.0955  &  1.321  &  0.0567  &  1.262  &  0.0542  \\
00037592020  &  57881.3  &  1023.9  &  3.06  &  0.073  &  2.66  &  0.1003  &  1.4  &  0.0557  &  1.259  &  0.0501  \\
00037592021  &  57888.4  &  991.4  &  2.88  &  0.063  &  2.165  &  0.0906  &  1.283  &  0.0595  &  1.194  &  0.0554  \\
00037592022  &  57896.2  &  581.9  &  2.61  &  0.092  &  1.027  &  0.0823  &  0.603  &  0.0538  &  0.345  &  0.0308  \\
00037592023  &  57902.4  &  946.5  &  2.78  &  0.063  &  1.887  &  0.0853  &  1.084  &  0.052  &  0.763  &  0.0366  \\
00037592024  &  57909.1  &  988.9  &  2.94  &  0.063  &  1.789  &  0.0877  &  1.022  &  0.0554  &  0.896  &  0.0486  \\
00037592025  &  57916.1  &  946.5  &  3.2  &  0.073  &  0.929  &  0.0609  &  0.542  &  0.0397  &  0.647  &  0.0474  \\
00037592026  &  57923.0  &  981.4  &  2.84  &  0.063  &  1.678  &  0.0885  &  0.763  &  0.0418  &  0.782  &  0.0429  \\
00037592027  &  57930.1  &  983.9  &  2.93  &  0.063  &  1.467  &  0.077  &  0.753  &  0.0419  &  0.73  &  0.0406  \\
00037592028  &  57937.8  &  1006.4  &  2.72  &  0.063  &  1.942  &  0.0837  &  1.084  &  0.05  &  0.807  &  0.0372  \\
00037592030  &  57951.1  &  874.1  &  3.26  &  0.073  &  2.273  &  0.1006  &  1.312  &  0.0652  &  1.406  &  0.0699  \\
00037592031  &  57958.3  &  966.5  &  3.03  &  0.073  &  1.896  &  0.0808  &  1.358  &  0.0661  &  0.506  &  0.0247  \\
00037592033  &  57968.0  &  946.5  &  3.67  &  0.082  &  2.796  &  0.1073  &  1.446  &  0.0597  &  1.544  &  0.0637  \\
00037592034  &  57972.5  &  978.9  &  3.68  &  0.082  &  2.746  &  0.1016  &  1.543  &  0.0623  &  1.466  &  0.0592  \\
00037592035  &  57979.1  &  998.9  &  3.4  &  0.073  &  1.506  &  0.0764  &  0.846  &  0.0463  &  1.052  &  0.0575  \\
00037592036  &  57986.9  &  926.5  &  3.82  &  0.092  &  1.631  &  0.0863  &  0.798  &  0.0441  &  0.822  &  0.0455  \\
00037592038  &  58000.1  &  759.2  &  3.83  &  0.092  &  1.207  &  0.0961  &  0.499  &  0.0396  &  1.145  &  0.0909  \\
00037592039  &  58007.3  &  949.0  &  4.81  &  0.104  &  2.101  &  0.0972  &  1.029  &  0.0503  &  1.278  &  0.0625  \\
00037592040  &  58014.2  &  1046.4  &  4.13  &  0.095  &  1.156  &  0.1103  &  0.339  &  0.0288  &  1.3  &  0.1103  \\
00037592041  &  58021.4  &  1058.8  &  4.13  &  0.095  &  0.752  &  0.0731  &  0.318  &  0.028  &  0.96  &  0.0843  \\
00037592042  &  58028.8  &  334.6  &  3.95  &  0.092  &  1.546  &  0.2617  &  0.656  &  0.0786  &  2.011  &  0.2412  \\
00037592043  &  58031.6  &  616.8  &  3.89  &  0.092  &  1.506  &  0.0996  &  0.805  &  0.0573  &  0.932  &  0.0663  \\
00037592044  &  58035.5  &  996.4  &  3.51  &  0.082  &  1.62  &  0.0821  &  0.856  &  0.0461  &  0.902  &  0.0486  \\
00037592045  &  58042.2  &  978.9  &  3.16  &  0.073  &  1.671  &  0.099  &  0.864  &  0.0549  &  1.101  &  0.0699  \\
00037592046  &  58056.8  &  751.7  &  2.99  &  0.073  &  0.885  &  0.074  &  0.5  &  0.0437  &  0.6  &  0.0524  \\
00037592047  &  58063.9  &  981.4  &  2.99  &  0.073  &  1.681  &  0.085  &  0.85  &  0.0459  &  0.931  &  0.0503  \\
00037592048  &  58070.9  &  891.5  &  2.66  &  0.063  &  1.875  &  0.0888  &  1.063  &  0.0545  &  1.026  &  0.0526  \\
00037592049  &  58077.5  &  856.6  &  2.5  &  0.063  &  0.758  &  0.0793  &  0.321  &  0.0306  &  0.893  &  0.0853  \\
00037592050  &  58084.6  &  1046.4  &  2.39  &  0.054  &  1.173  &  0.0837  &  0.458  &  0.0322  &  1.355  &  0.0955  \\
00037592051  &  58090.2  &  973.9  &  2.86  &  0.063  &  2.574  &  0.1126  &  1.313  &  0.0594  &  1.847  &  0.0836  \\
00037592052  &  58098.0  &  869.1  &  2.82  &  0.063  &  2.031  &  0.1066  &  1.065  &  0.0612  &  1.393  &  0.08  \\
00037592056  &  58130.7  &  981.4  &  3.83  &  0.082  &  2.969  &  0.1143  &  1.691  &  0.0714  &  1.572  &  0.0664  \\
00037592057  &  58132.8  &  983.9  &  3.82  &  0.082  &  2.812  &  0.1  &  1.658  &  0.0636  &  1.125  &  0.0432  \\
00037592058  &  58134.3  &  1028.9  &  3.34  &  0.073  &  2.887  &  0.102  &  1.607  &  0.0617  &  1.507  &  0.0579  \\
00037592059  &  58136.8  &  794.1  &  3.7  &  0.082  &  2.828  &  0.1166  &  1.614  &  0.0727  &  1.463  &  0.0659  \\
00037592060  &  58138.5  &  993.9  &  3.78  &  0.082  &  2.126  &  0.0932  &  1.097  &  0.0517  &  1.056  &  0.0498  \\
00037592061  &  58140.6  &  876.5  &  3.87  &  0.082  &  2.57  &  0.1042  &  1.468  &  0.0647  &  1.098  &  0.0484  \\
00037592062  &  58142.8  &  1063.8  &  3.72  &  0.082  &  2.489  &  0.094  &  1.384  &  0.0561  &  1.153  &  0.0467  \\
00037592063  &  58144.8  &  956.5  &  3.68  &  0.082  &  2.549  &  0.1048  &  1.361  &  0.0602  &  1.186  &  0.0524  \\
00037592064  &  58146.8  &  976.4  &  3.34  &  0.073  &  2.941  &  0.1153  &  1.552  &  0.0652  &  1.453  &  0.0611  \\
00037592065  &  58148.9  &  1066.3  &  3.74  &  0.082  &  2.435  &  0.0902  &  1.436  &  0.0575  &  1.046  &  0.0419  \\
00037592066  &  58150.1  &  986.4  &  3.68  &  0.082  &  2.304  &  0.0907  &  1.386  &  0.06  &  1.018  &  0.0441  \\
00037592067  &  58152.0  &  751.7  &  3.88  &  0.082  &  2.158  &  0.108  &  1.16  &  0.0615  &  0.992  &  0.0526  \\
00037592068  &  58154.0  &  993.9  &  4.12  &  0.095  &  2.992  &  0.1063  &  1.719  &  0.0668  &  1.464  &  0.0568  \\
00037592069  &  58156.0  &  814.1  &  4.21  &  0.095  &  2.842  &  0.108  &  1.843  &  0.0763  &  0.979  &  0.0406  \\
00037592070  &  58158.2  &  1266.1  &  4.29  &  0.104  &  2.329  &  0.0812  &  1.419  &  0.0539  &  1.061  &  0.0403  \\
00037592071  &  58160.6  &  1071.3  &  4.3  &  0.095  &  2.22  &  0.0934  &  1.166  &  0.0519  &  1.227  &  0.0546  \\
00037592072  &  58162.9  &  1026.4  &  4.08  &  0.095  &  1.646  &  0.0735  &  1.036  &  0.052  &  0.776  &  0.039  \\
00037592073  &  58164.0  &  971.4  &  4.23  &  0.095  &  2.194  &  0.0901  &  1.279  &  0.0577  &  1.046  &  0.0472  \\
00037592074  &  58166.7  &  414.6  &  4.35  &  0.104  &  2.195  &  0.1395  &  1.336  &  0.0949  &  1.171  &  0.0831  \\
00037592075  &  58168.0  &  956.5  &  4.34  &  0.095  &  2.053  &  0.0857  &  1.289  &  0.0626  &  1.041  &  0.0506  \\
00037592076  &  58170.1  &  968.9  &  4.49  &  0.104  &  2.672  &  0.1102  &  1.463  &  0.0653  &  1.218  &  0.0544  \\
00037592077  &  58172.2  &  866.6  &  4.61  &  0.104  &  4.008  &  0.207  &  2.271  &  0.1237  &  1.812  &  0.0987  \\
00037592078  &  58174.1  &  981.4  &  4.52  &  0.104  &  3.313  &  0.1369  &  1.616  &  0.0705  &  2.012  &  0.0877  \\
00037592079  &  58176.7  &  1056.4  &  4.51  &  0.104  &  2.645  &  0.0957  &  1.562  &  0.0608  &  1.136  &  0.0442  \\
00037592080  &  58178.7  &  981.4  &  4.24  &  0.095  &  1.641  &  0.0872  &  0.775  &  0.0431  &  0.928  &  0.0516  \\
00037592081  &  58180.4  &  799.1  &  4.53  &  0.104  &  2.605  &  0.1077  &  1.591  &  0.0729  &  1.196  &  0.0548  \\
00037592082  &  58182.1  &  1013.9  &  4.72  &  0.104  &  1.095  &  0.0614  &  0.698  &  0.0427  &  0.617  &  0.0378  \\
00037592083  &  58184.5  &  1006.4  &  4.79  &  0.104  &  2.02  &  0.1279  &  1.391  &  0.0977  &  1.573  &  0.1105  \\
00037592084  &  58186.3  &  1006.4  &  4.51  &  0.104  &  1.512  &  0.0749  &  0.909  &  0.0483  &  0.671  &  0.0356  \\
00037592085  &  58188.2  &  939.0  &  4.66  &  0.104  &  2.246  &  0.0954  &  1.336  &  0.0636  &  1.318  &  0.0627  \\
00037592086  &  58190.4  &  1026.4  &  4.31  &  0.095  &  1.541  &  0.0792  &  0.784  &  0.0427  &  0.934  &  0.0509  \\
00037592087  &  58192.6  &  901.5  &  4.24  &  0.104  &  2.064  &  0.0974  &  1.14  &  0.0576  &  1.057  &  0.0534  \\
00037592088  &  58194.8  &  998.9  &  4.21  &  0.095  &  1.139  &  0.0665  &  0.643  &  0.0426  &  1.056  &  0.07  \\
00037592089  &  58196.8  &  1013.9  &  4.38  &  0.104  &  0.746  &  0.0617  &  0.33  &  0.0285  &  0.694  &  0.0599  \\
00037592090  &  58198.4  &  866.6  &  4.42  &  0.104  &  1.001  &  0.0758  &  0.487  &  0.0383  &  0.998  &  0.0784  \\
00037592091  &  58200.6  &  1088.8  &  4.34  &  0.095  &  0.677  &  0.0997  &  0.259  &  0.0336  &  0.838  &  0.1089  \\
00037592092  &  58202.8  &  1018.9  &  4.41  &  0.104  &  0.374  &  0.0425  &  0.171  &  0.0216  &  0.421  &  0.0534  \\
00037592093  &  58204.7  &  601.9  &  4.63  &  0.104  &  0.345  &  0.0499  &  0.237  &  0.0383  &  0.239  &  0.0386  \\
00037592094  &  58206.8  &  956.5  &  4.8  &  0.104  &  0.492  &  0.0586  &  0.403  &  0.0606  &  1.852  &  0.2784  \\
00037592095  &  58208.9  &  292.2  &  5.11  &  0.124  &  1.226  &  0.1468  &  0.523  &  0.0657  &  0.859  &  0.1078  \\
00037592098  &  58215.0  &  896.5  &  5.58  &  0.136  &  2.446  &  0.1035  &  1.565  &  0.0739  &  1.145  &  0.054  \\
00037592099  &  58216.7  &  664.3  &  5.05  &  0.114  &  2.04  &  0.1032  &  1.387  &  0.078  &  0.564  &  0.0317  \\
00037592100  &  58218.3  &  836.6  &  5.51  &  0.124  &  3.107  &  0.1143  &  1.863  &  0.0766  &  1.643  &  0.0675  \\
00037592101  &  58222.7  &  1106.3  &  5.19  &  0.114  &  2.577  &  0.0951  &  1.497  &  0.0613  &  1.536  &  0.0629  \\
00037592103  &  58226.8  &  1106.3  &  4.96  &  0.114  &  2.058  &  0.0845  &  1.1  &  0.0488  &  0.981  &  0.0436  \\
00037592104  &  58228.4  &  1311.1  &  4.44  &  0.104  &  1.827  &  0.075  &  1.006  &  0.0448  &  0.85  &  0.0379

\end{longtable}


\begin{references}

\reference{} Bentz, M. C., \& Katz, S. 2015, PASP, 127, 67

\reference{} Buisson, D. J. K., et al. 2018, MNRAS, 475, 2306

\reference{} Burrows, D. N., et al. 2005, Sp. Sci. Rev., 120, 165

\reference{} Cackett, E. M., et al. 2018, ApJ, 857, 53

\reference{} Denney, K. D., Peterson, B. M., Pogge, R. W., et al. 2010, ApJ, 721, 715

\reference{} Dickey, J. M., \& Lockman, F. J. 1990, ARAA, 28, 215

\reference{} Edelson, R., Gelbord, J., Cackett, E., et al. 2017, ApJ, 840, 41

\reference{} Edelson, R., Gelbord, J. M., Horne, K., et al. 2015, ApJ, 806, 129

\reference{} Edelson, R. A., \& Krolik, J. H. 1988, ApJ, 333, 646

\reference{} Gallo, L. C., et al. 2018, MNRAS, 478, 2557

\reference{} Gardner, E., \& Done, C. 2017, MNRAS, 470, 3591

\reference{} Horne, K., Peterson, B. M., Collier, S. J., \& Netzer, H., PASP, 2004, 116, 465

\reference{} Kaastra, J., Kriss, G., Cappi, M., et al., 2014, Science, 345, 64

\reference{} Korista, K. T., \& Goad, M. R. 2001, ApJ, 553, 695

\reference{} Markoff, S., Nowak, M. A., \& Wilms, J. 2005, ApJ, 635, 1203

\reference{} McHardy, I. M., et al. 2014, MNRAS, 444, 1469

\reference{} McHardy, I. M., et al. 2018, MNRAS, 480, 2881

\reference{} Miller, J. M., Cackett, E., Zoghbi, A., et al. 2018, ApJ, 865, 97

\reference{} Miller, J. M., Homan, J., Steeghs, D., et al. 2006, ApJ, 653, 525

\reference{} Pal, M., \& Naik, S. 2018, MNRAS, 474, 5351

\reference{} Peterson, B. M., Wanders, I., Bertram, R., et al. 1998, ApJ, 501, 82

\reference{} Poole, T. S., et al. 2008, MNRAS, 383, 627

\reference{} Robertson, D. R. S., et al. 2015, MNRAS, 453, 3455

\reference{} Roming, P. W. A., et al. 2005, Sp. Sci. Rev., 120, 95

\reference{} Starkey, D., Horne, K., Fausnaugh, M. M., et al. 2017, ApJ 835, 65

\reference{} Timmer, J., \& K\"onig M. 1995, A\&A, 300, 707

\reference{} Troyer, J., et al. 2016, MNRAS, 456, 4040

\reference{} Uttley, P., Edelson, R., McHardy, I. M., et al. 2003, ApJL, 584, L53

\reference{} Vaughan, S., et al. 2003, MNRAS, 345, 1271

\reference{} Welsh, W. F. 1999, PASP, 111, 1347

\reference{} Winter, L. M., Danforth, C., Vasudevan, R., et al. 2011, ApJ, 728, 28

\reference{} Woo, J.-H., \& Urry, C. M. 2002, ApJ, 579, 530

\reference{} Zoghbi, A., Miller, J. M., Cackett, E. M., et al., 2018, ApJ, in prep.

\reference{} Zoghbi, A., Reynolds, C., \& Cackett, E. M. 2013, ApJ, 777, 24

\end{references}
\end{document}